\documentstyle{article}
\begin{document}
\title{
Specific heat of the Coulomb glass
}
\author{ By
A.~M\"obius\\
Institute for Solid State and Materials Research Dresden,\\
D-01171 Dresden, Germany,\\
\\
P.~Thomas\\ 
Department of Physics and Material Sciences Center,\\
Philipps University, Renthof 5, D-35032 Marburg, Germany,\\
\\
J.~Talamantes\\
Department of Physics, California State University,\\
Bakersfield, CA 93311, USA,\\ 
\\
and C.J.~Adkins\\
Cavendish Laboratory,\\ 
Madingley Road, Cambridge CB3 0HE, UK\\
}

\date{April 30, 2001}

\maketitle

\begin{abstract}
The specific heat $c$ of the Coulomb glass is studied by numerical
simulations. Both the lattice model with various strengths of 
disorder, and the random-position model are considered for the one- to
three-dimensional cases. In order to extend the investigations down 
to very low temperatures where the many-valley structure of the 
configuration space is of great importance we use a 
hybrid-Metropolis procedure. This algorithm bridges the gap between 
Metropolis simulation and analytical statistical mechanics. The 
analysis of the simulation results shows that the correlation length
of the relevant processes is rather small, and that multi-particle 
processes yield an essential contribution to the specific heat in 
all cases except the one-dimensional random-position model.
\end{abstract}

\section{Introduction}

The Coulomb glass, a prominent model of disordered systems of 
localised particles with long-range interaction, has been extensively
studied for almost three decades. It plays an important role as a
semi-classical model for dilute impurity bands when quantum 
interference as well as thermal excitation to the conduction band or 
from the valence band can be neglected. Various aspects of this 
model have been investigated since the pioneering works by Pollak
(1970), Srinivasan (1970), Efros and Shklovskii (1975), and Kurosawa 
and Sugimoto (1975); for related reviews see Shklovskii and Efros
(1984), Efros and Shklovskii (1985), Pollak and Ortu\~{n}o (1985),
and Ortu\~{n}o {\sl et al}.\ (2001).  Nevertheless, this area of 
research has not lost its attraction, as is shown, for example, by 
recent studies of aging effects in electric conduction (Vaknin 
{\sl et al}.\ 2000), of the possibility of a phase transition 
(Grannan and Yu 1993, Vojta and Schreiber 1994, D\'\i az-S\'anchez 
{\sl et al}.\ 2000b), and by combined investigations of dc 
conductivity and tunnel conductance (Sandow {\sl et al}.\ 2001).

An important and characteristic feature of the Coulomb glass is the 
many-valley structure of its configuration space, not only concerning 
single-particle hops, but also with respect to some classes of more
complex excitations. In this respect it resembles the Ising spin 
glass and other disordered systems, see e.g.\ M\'ezard {\sl et al}.\
(1987). This configuration-space structure is related to the existence
of a huge number of so-called pseudo-ground states, as was observed 
in the first numerical Coulomb glass simulations (Baranovskii 
{\sl et al}.\ 1976). The spectrum of these pseudo-ground states has 
recently been investigated in detail by Kogan (1998). Moreover, the 
related importance of many-particle excitations is directly 
illustrated by Fig.\ 3 in Schreiber and Tenelsen (1993). In
particular, the huge number of local minima has been shown to cause a 
glass like relaxation behaviour (Schreiber {\sl et al}.\ 1996,
P\'erez-Garrido {\sl et al}.\ 1999) and non-ergodic effects 
(D\'\i az-S\'anchez {\sl et al}.\ 2000a).

However, under certain conditions, also ``equilibrium'' observables 
can reflect the many-valley character of the configuration space. We 
focus here on the specific heat $c$ (heat capacity per site), in 
particular on its temperature dependence $c(T)$ for various dimensions
$d$. It has been studied numerically by several groups (Baranovskii 
{\sl et al}.\ 1982, Davies {\sl et al}.\ 1982, Davies {\sl et al}.\ 
1984, Schreiber and Tenelsen 1993, Tenelsen and Schreiber 1994, 
M\"obius and Pollak 1996, M\"obius and Thomas 1997), mainly by means 
of the Metropolis procedure (Metropolis {\sl et al}.\ 1953). However, 
this approach is inappropriate in the low-temperature region.  One of 
the reasons for its failure is a direct consequence of the many-valley 
structure of the configuration space: the simulations usually get 
stuck in some local minimum. Consequently, other methods have to be 
used. Thus, almost complete sets of system states were obtained for 
the direct calculation of the low-$T$ behaviour of $c(T)$ by means of 
different algorithms in Schreiber and Tenelsen (1993), Tenelsen and 
Schreiber (1994), M\"obius and Pollak (1996), M\"obius and Thomas 
(1997), compare also Mochena and Pollak (1991) and P\'erez-Garrido 
{\sl et al}.\ (1997). Unfortunately, these specific heat studies have 
so far been restricted to the three-dimensional lattice model. 

To obtain an analytical description of $c(T)$, Baranovskii 
{\sl et al.}\ (1980) considered the excitation from the ground 
state by independent single-particle hops. In the case of randomly 
positioned sites, the adequacy of this approach was supported by 
numerical experiments (Baranovskii {\sl et al.}\ 1982). However, for 
the three-dimensional lattice case, M\"obius and Pollak (1996) showed
(though only for one value of the disorder strength) that this 
description considerably underestimates $c(T)$.

The aim of our present work is to study the temperature dependence of
the specific heat $c(T)$ in the general case, considering in 
particular also that $T$ region which is not accessible to usual 
Metropolis simulations. We have simulated one- to three-dimensional 
samples with both random and lattice structures, and, in the latter 
case, we have varied the strength of the disorder. To illuminate the 
puzzling situation described in the previous paragraph, we have sought 
for the conditions under which $c(T)$ can be understood in terms of 
independent single-particle hops. For that purposes, the computer code 
which was developed in M\"obius and Thomas (1997) for the 
three-dimensional lattice case has been rewritten and improved to 
handle arbitrary structures.

\section{Model}

All situations which are to be studied can be described by the
Hamiltonian 
\begin{equation}
H = \sum_\alpha \varepsilon_\alpha \, n_\alpha +
\frac{1}{2} \sum_{\alpha \neq \beta}
\frac{(n_\alpha - K)\,(n_\beta - K)}
{\left|\vec{x}_\alpha-\vec{x}_\beta\right|}\,
\end{equation}  
where the first sum represents an on-site disorder contribution and
the second sum describes the Coulomb interaction, see e.g.\ 
Shklovskii and Efros (1984). The sums are over the $N$ sites of the 
sample. These sites positioned at $\vec{x}_\alpha$ are arranged either 
on a lattice (lattice model), or randomly (random-position model), the 
more realistic case. The occupation number $n_\alpha \in \{1,0\}$ 
indicates whether or not site $\alpha$ is occupied by one of the 
mobile charges -- note that, in a donor band, $n_\alpha = 1$ 
corresponds to a missing electron. In the lattice model, the 
$\varepsilon_\alpha$ describe only the on-site disorder potential; 
they are uniformly distributed between $-B/2$ and $B/2$ where the 
parameter $B$ characterises the strength of the disorder. In this 
case, for electroneutrality, background charges $-K$ are attached to 
each site where $K = \langle n_\alpha \rangle$. However, for an n-type
impurity band emulated by means of the random-position model, the 
$\varepsilon_\alpha$ denote the potentials of the charged acceptors 
and we have $K = 0$. Elementary charge, spatial density of the 
$\vec{x}_\alpha$, and dielectric constant are all taken to be 1.

In order to reduce finite-size effects and to exclude the influence
of surfaces, we have modified the interaction term in our simulations
by imposing periodic boundary conditions and using the minimum image
convention (Metropolis {\sl et al}.\ 1953, Davies {\sl et al}.\ 1982).
Thus our sample is an elementary cell in a (super)lattice, and the 
interaction energy of two sites is assumed to be determined by the 
shortest distance within this repeated representation. Moreover, we 
take into account particle exchange with a reservoir. That means we 
consider the grand canonical potential 
$h = H - \mu \sum_\alpha n_\alpha$, where the chemical potential 
$\mu$, if it is not already fixed by symmetry reasons, is determined 
by relaxing the closed system before starting the Metropolis part of 
the simulation. Considering the grand canonical ensemble rather than 
the canonical has several advantages. Correlation times are reduced by
emulating in this manner also long-range hops. Moreover, in particular
at high $T$, the influence of finite-size effects is diminished. 
However, one must remember that any $T$-dependence of $\mu$ may 
influence $c(T)$, as it does in metals. We have taken care to exclude 
such effects in the tests of our code.

\section{Algorithm}

In order to treat correlations completely, we have studied
the specific heat of the Coulomb glass by numerical simulations. The 
standard procedure for this is the Metropolis algorithm (Metropolis 
{\sl et al}.\ 1953) which consists of substituting time averaging for
ensemble averaging. For related surveys see e.g.\ Binder (1984) or
Binder (1997), and references therein. However, in this approach, at 
low $T$ and close to phase transitions, large difficulties arise from
diverging correlation times. In the low-$T$ case, these difficulties 
are reduced to a large extent by the hybrid algorithm developed in 
M\"obius and Thomas (1997). This bridges the gap between Metropolis 
simulation and analytical statistical mechanics. In the present work,
we use an improved version of this procedure.

The difficulties mentioned above are tackled here in two ways 
following M\"obius and Thomas (1997). First, in order to handle the 
problem of the great portion of unsuccessful attempts at low 
$T$, we do not perform random trials of state modifications at 
discrete time values as in the original Metropolis approach. Rather, 
on a continuous time scale, we focus on the state modifications
which are really performed, though each such state investigation is 
clearly more elaborate than the trials in the standard procedure: To 
obtain the mean dwell time at the current state, the transition 
probabilities to all of its neighbours in the configuration space are 
calculated; the next state (in time) is picked randomly according to 
the ratios of the various transition probabilities, see M\"obius and 
Thomas (1997). This approach resembles the n-fold way algorithm 
(Bortz {\sl et al}.\ 1975). It becomes increasingly advantageous in 
comparison to the standard Metropolis procedure as $T$ is lowered. 
Moreover, problems which can arise from time discretization (Adam
{\sl et al}.\ 1999) are avoided in this way.

Second, our hybrid algorithm takes advantage of the fact that 
thermodynamic averages are independent of the dynamics considered 
provided only that detailed balance is guaranteed. Thus, in order to 
reduce the problems arising from the huge number of local minima, it 
decomposes the configuration space into low- and high-energy parts, 
and utilises very fast auxiliary transitions additionally introduced 
between the states in the low-energy part. Therefore, the sample 
simulated thermalises very rapidly with respect to the low-energy part
of the configuration space as soon as its state enters this region: 
Due to the fast auxiliary transitions, the dwell time in this 
low-energy part as a whole is large in comparison to the dwell time in
one of the states contained in it. Thus all quantities can be assumed 
to have the corresponding thermal averages of the low-energy subset of
the configuration space while the sample stays in this configuration 
space region. Moreover, the escape probability from this subset does 
not depend on the entrance state. For details we have to refer to 
M\"obius and Thomas (1997). For related approaches see Frantz 
{\sl et al}.\ (1990) and Krauth and Pluchery (1994).

In the present work, we have modified these ideas in three ways. We 
have used an improved algorithm for obtaining an almost complete set 
of low-energy system states. In M\"obius and Thomas (1997), this 
subset was constructed by first sampling local minima, and then 
stepwise improving this subset by adding all the low-energy states 
found during the simulation, see also Schreiber and Tenelsen (1993). 
Here, we have extended the subset construction as introduced in 
D\'\i az-S\'anchez {\sl et al}.\ (2000a). That means, we first sample
local minima; then we improve this backbone of the low-energy subset 
by thermal cycling (M\"obius {\sl et al}.\ 1997) (one could also 
compare this optimisation procedure with compaction of a granular
medium, such as ground coffee, by shaking). Finally, we search 
systematically the surroundings of all low-energy system states that 
have been found so far for further states (M\"obius and Pollak 1996,
P\'erez-Garrido {\sl et al}.\ 1997). Also in our new code, we add the
states visited during the Metropolis simulation to the low-energy 
states set if their energy is lower than the highest energy value in 
this subset. That this happens very seldom in practice, confirms the 
efficiency of the three-stage low-energy subset construction 
procedure, described above.
 
As has often been stressed by Mike Pollak, see e.g.\ the review 
Pollak and Ortu\~{n}o (1985), cluster excitations are expected to play
an essential role in the low-$T$ dynamics of the Coulomb glass. In a 
direct way, one can include only small, compact clusters into the 
Metropolis dynamics. Unfortunately, due to the long-range character of
the interaction, the usual cluster algorithms as proposed in 
Swendsen and Wang (1987) seem not to be applicable. Therefore, we use 
another approach here: Comparing repeatedly small numbers of states 
from the low-energy subset (mainly picked randomly), we search for 
groups of sites which change occupation only simultaneously, see 
Kawashima and Suzuki (1992). These clusters are tabulated and included
into the simulation dynamics. 

Finally, in order to prevent the low-$T$ improvements, described 
above, from slowing down too much the computer code in the high-$T$ 
case, we have introduced a $T$ dependent neighbourhood pruning: The 
number of the transitions included in the Metropolis dynamics is 
appropriately reduced with increasing temperature.

\section{Numerical results}

A series of tests has been performed to check and optimise the new 
code. The quality of the sets of low-energy states was confirmed 
for various degrees of disorder by series of runs in which 
we treated the same sample repeatedly. Moreover, we simulated lattice
samples without disorder and reproduced the ordered ground states. In
these tests, when we considered half-filled three-dimensional samples
of $(2m)^3$ sites ($m$ integer), we observed the NaCl structure to be
the ground state only for odd $m$. The six-fold degenerate layered 
structure has slightly lower energies for arbitrary even $m$. At
present, we cannot answer the question whether this strange feature 
of the long-range Coulomb interaction is only an artifact of the 
boundary conditions, or  whether it has some physical meaning. 
However, it is remarkable in this context that layered structures 
have been observed in quench simulations of one-component plasmas
(Ogata and Ichimaru 1989). Additionally, we have considered lattice 
samples without disorder where only a fraction $2^{-p}$ of the sites 
is occupied ($p$ integer), studied analytically in Hunt and Pollak 
(1986). For $d = 3$, $p = 2$, and $m = 4$, we obtained ground states 
with layered structure.

We turn now to the discussion of the $c(T)$ data obtained in the 
simulations. We have compared our results on the random-position 
model with those obtained previously by Baranovskii {\sl et al}.\ 
(1982): we observed very good agreement except for slightly larger 
peak values in our case. (In the present study, the smallest 
considered $T$ value is lower than in Baranovskii {\sl et al}.\ (1982)
by a factor 16.) Furthermore, also the comparison with lattice data 
from M\"obius and Thomas (1997) showed perfect agreement within the
statistical accuracy.

Finite-size effects can considerably complicate Coulomb glass 
simulations due to the long-range character of the interaction. 
Such problems are well-known from the study of the single-particle 
density of states where the sample size is related to a cut-off 
energy of the interaction spectrum and thus to a reliability limit 
in the energy dependence of this density of states (Baranovskii 
{\sl et al}.\ 1979, M\"obius {\sl et al}.\ 1992). In consequence, the
thermal smearing of the Coulomb gap is connected with a correlation 
length diverging as $T \rightarrow 0$ (Hunt 1990). However, those 
correlations and corresponding processes which are essential for 
the formation of the Coulomb gap need not also be the most important 
for the specific heat. Other state modifications rather than 
single-particle hops can be far more important for this quantity. 
This assertion is suggested by the numerical data from 
M\"obius and Pollak (1996) showing that, for $d = 3$, $c(T)$ data seem
to be reliable down to $T$ values far smaller than the cut-off energy 
of the interaction spectrum.

To illuminate this point, we have plotted $c(T)$ for various sample 
sizes in Fig.\ 1, considering the lattice model with fixed disorder 
strength with $d = 1$, 2, and 3. For $d = 1$, the curves for $N = 12$,
and 400 sites agree perfectly (within the statistical accuracy of our
data) above as well as below the cut-off of the interaction energies, 
$2/N$, of the smaller sample. On the other hand, considerable 
finite-size effects are present for $N = 3$ over the range $T < 1$.
Thus the specific heat is mainly determined by short-range single- or 
multi-particle hops here. There are large $c(T)$ fluctuations in the 
smaller samples at the lowest $T$ values (Tenelsen and Schreiber 1994,
M\"obius and Pollak 1996). These fluctuations seem to be ``localised''
in the big samples.

In the case $d = 2$, significant finite-size effects are present for 
$N = 9$ and $N = 25$. However, the deviation between the data for 
samples of 100 and 900 sites is very small, just at the accuracy 
limit. For $d = 3$, finite-size effects are clearly larger than in
the previous cases. However, the results for samples of 512 and 1000
sites agree well with each other. 

The convergence properties shown in Fig.\ 1 yield a rough estimate 
of the correlation radius: $l_{\rm corr} = 3...5$. This length seems
to increase ``slowly'' with decreasing $T$. 

Fig.\ 1 shows that the finite-size effects, which arise from the 
Coulomb interaction, rise with increasing $d$. To illuminate the 
interplay of interaction and dimension directly, Fig.\ 2 compares 
$c(T)$ data for various $d$ from lattice simulations with fixed 
disorder strength with the interaction-free case. In the latter 
situation, analytical calculations yield that $c \propto T$ for 
$T \rightarrow 0$ and that  $c \propto T^{-2}$ if 
$T \rightarrow \infty$, in agreement with the numerical data. 
Switching on the Coulomb interaction causes $c$ to decrease at low 
$T$ and to increase at high $T$, the more the larger $d$ is. In the 
low-$T$ region, $c(T)$ decreases faster than linear as 
$T\rightarrow 0$. Though $c(T)$ cannot be approximated by a simple 
power law in the low-$T$ region studied, our data raises the question
of whether $c(T)$ might tend to a linear dependence again if $T$ is 
sufficiently lowered.

The influence of disorder strength is studied in Figs.\ 3 and 4,
showing $\log_{10} c$ and $c$, respectively, in dependence on 
$\log_{10} T$. The observation that, for high $T$, increasing $B$ 
reflects as some shift of $c(T)$ towards higher $T$, which was 
reported for $d =3$ in M\"obius and Thomas (1997), is made here also 
for $d = 2$ and $d = 1$.  However, at low $T$, there is a weaker, but 
more complicated dependence on the disorder strength. In all cases, 
the random-position model yields broader $c(T)$ curves than the 
lattice model. At low $T$, these data are close to the results 
obtained for low disorder strength in the lattice model.

A particular feature of $c(T)$ is only seen clearly when $c$ is 
plotted on a linear scale as in Fig.\ 4: In the lattice model, for
$B = 0.5$ and $d = 2$ as well as $d = 3$, the $c(T)$ curve is narrower 
and considerably higher than for the other $B$ values. The question, 
whether or not this behaviour might be a precursor of a phase 
transition for still lower $B$ (Grannan and Yu 1993, Vojta and 
Schreiber 1994, D\'\i az-S\'anchez {\sl et al}.\ 2000b) cannot be 
answered yet. Corresponding simulations are extremely CPU-time 
consuming.

Finally, we address the question of which processes determine $c(T)$ 
as $T \rightarrow 0$. To clarify which contributions to the specific 
heat yield single-particle hops (SPH) and multi-particle hops, 
respectively, we have calculated the value $c_{\rm SPH}$, which one 
obtains when considering SPHs to be the only relevant, independent 
excitations of the ground state. The $T$-dependence of the quotient of
$c_{\rm SPH}$ and of the ``exact'' $c(T)$ data is represented in Fig.\ 
5 for various $d$ and $B$, for the lattice as well as for the 
random-position model. The SPH approximation obviously begins to fail 
when $T$ is sufficiently increased for a significant number of sites 
to belong simultaneously to more than one of the excited SPHs. 
Therefore, Fig.\ 5 focuses on that region, where the corresponding 
overlap correction does not exceed $0.1\,c$; quotient values for 
higher $T$, included in certain cases only, show that the breakdown of
the above condition is correlated with a qualitative change of the 
quotient's $T$ dependence. The maximum applicability $T$ of the SPH 
approximation decreases with increasing sample size. Therefore, for 
$d = 1$ and 2, medium sample sizes are considered here. Nevertheless, 
quotient values could only be obtained for a $T$ region which cannot 
be studied by standard Metropolis simulations, and, for the 
random-position model in the case $d = 1$, the applicability region 
of the SPH approximation includes only one data point.

Fig.\ 5 shows that the SPH approximation, when applicable, clearly 
underestimates $c$ for all but one case; the random-position model
for $d = 1$ is the only exception. In more detail, in the two- and 
three-dimensional lattice models, multi-particle contributions are
particularly important for weak disorder. In the random-position
model, the relevance of multi-particle excitations increases with
the dimension. 

Thus, Fig.\ 5 implies the conclusion that, in agreement with Mike
Pollak's view on the Coulomb glass model (Pollak and Ortu\~{n}o 1985),
multi-particle processes yield an essential contribution to the
specific heat in the low-temperature limit. 

\vspace{0.5cm}

\begin{center}
{\sl Acknowledgments}
\end{center}

Part of this study was performed during a stay of AM at the 
California State University, Bakersfield. He thanks the US National
Science Foundation for financing this visit under grant DMR-9803686, 
and the colleagues in Bakersfield for their hospitality. Moreover,
we are much obliged to A. Hunt for critical discussions, and to 
U.~Kleinekath\"ofer, S.~K\"ostlmeier, and M.~Schreiber for the
possibility to perform part of the simulations at workstations
of the Technische Universit\"at Chemnitz, and for various related 
help.

\newpage

\begin{center}
{\sl References}
\end{center}

\begin{description}

\item Adam, E., Billard, L., and Lan\c{c}on, F., 1999, {\sl Phys.\ 
Rev.\ E}, {\bf 59}, 1\,212.

\item Baranovskii, S.D., Efros, A.L., Gelmont, B.L., and Shklovskii, 
B.I., 1976, {\sl J.\ Phys.\ C: Solid State Phys.}, {\bf 9}, 2021. 

\item Baranovskii, S.D., Efros, A.L., Gelmont, B.L., and Shklovskii, 
B.I., 1979, {\sl J.\ Phys.\ C: Solid State Phys.}, {\bf 12}, 1023.

\item Baranovskii, S.D., Shklovskii, B.I., and Efros, A.L., 1980, 
{\sl Zh.\ Eksp.\ Teor.\ Fiz.}, {\bf 78}, 395 [Transl.: 1980, 
{\sl Sov.\ Phys.\ JETP}, {\bf 51}, 199]. 

\item Baranovskii, S.D., Uzakov, A.A., and Efros, A.L., 1982, 
{\sl Zh.\ Eksp.\ Teor.\ Fiz.}, {\bf 83}, 756 [Transl.: 1982, 
{\sl Sov.\ Phys.\ JETP}, {\bf 56}, 422].

\item Binder, K., editor, 1984, {\sl Applications of the Monte 
Carlo Method in Statistical Physics}, in {\sl Topics in Current 
Physics}, Vol.\ 36, (Berlin: Springer).

\item Binder, K., 1997, {\sl Rep.\ Prog.\ Phys.}, {\bf 60}, 487.

\item Bortz, A.B., Kalos, M.H., and Lebowitz, J.L., 1975, {\sl J.\ 
Comp.\ Phys.}, {\bf 17}, 10.

\item Davies, J.H., Lee, P.A., and Rice, T.M., 1982, {\sl Phys.\ Rev.\
Lett.}, {\bf 49}, 758.

\item Davies, J.H., Lee, P.A., and Rice, T.M., 1984, {\sl Phys.\ Rev.\ 
B}, {\bf 29}, 4260. 

\item D\'\i az-S\'anchez, A., M\"obius, A., Ortu\~no, M., Neklioudov, 
A., and Schreiber, M., 2000a, {\sl Phys.\ Rev.\ B}, {\bf 62}, 8\,030.

\item D\'\i az-S\'anchez, A., Ortu\~no, M., P\'erez-Garrido, A., and 
Cuevas, E., 2000b, {\sl phys.\ stat.\ sol.}, (b), {\bf 218}, 11.

\item Efros, A.L., and Shklovskii, B.I., 1975, {\sl J.\ Phys.\ C}, 
{\bf 8}, L49. 

\item Efros, A.L., and Shklovskii, B.I., 1985, {\sl Electron-electron 
interactions in disordered systems}, edited by Efros, A.L., and 
Pollak, M., (Amsterdam - Oxford - New York - Tokyo: North Holland), 
p.\ 409.

\item Frantz, D.D., Freeman, D.L., and Doll, J.D., 1990, {\sl J.\ 
Chem.\ Phys.}, {\bf 93}, 2769.

\item Grannan, E.R., and Yu, C.C., 1993, {\sl Phys.\ Rev.\ Lett.}, 
{\bf 71}, 3\,335.

\item Hunt, A., and Pollak, M., 1986, {\sl Phil.\ Mag.\ B}, {\bf 53}, 
353.

\item Hunt, A., 1990, {\sl Phil.\ Mag.\ Lett.}, {\bf 62}, 371.

\item Kawashima, N., and Suzuki, M., 1992, {\sl J.\ Phys.\ A: Math.\
Gen.}, {\bf 25}, 1055.

\item Kogan, S., 1998, {\sl Phys.\ Rev.\ B}, {\bf 57}, 9\,736.

\item Krauth, W., and Pluchery, O., 1994, {\sl J.\ Phys.\ A: Math.\ 
Gen.}, {\bf 27}, L715.

\item Kurosowa, T., and Sugimoto, H., 1975, {\sl Progr.\ Theor.\ 
Phys.\ Suppl.}, {\bf 57}, 217.

\item Metropolis, N., Rosenbluth, A.W., Rosenbluth, M.N., Teller, 
A.H., and Teller, E., 1953, {\sl J.\ Chem.\ Phys.}, {\bf 21}, 1087.

\item M\'ezard, M., Parisi, G., and Virasoro, M.A., editors, 1987, 
{\sl Spin glass theory and beyond}, (Singapore: World Scientific).  

\item M\"obius, A., Richter, M., and Drittler, B., 1992, {\sl Phys.\ 
Rev.\ B}, {\bf 45}, 11\,568.

\item M\"obius, A., and Pollak, M., 1996, {\sl Phys.\ Rev.\ B}, 
{\bf 53}, 16\,197.

\item M\"obius, A., and Thomas, P., 1997, {\sl Phys.\ Rev.\ B}, 
{\bf 55}, 7\,460.

\item M\"obius, A., Neklioudov, A., D\'\i az-S\'anchez, A.,
Hoffmann, K.H., Fachat, A., and Schreiber, M., 1997, {\sl Phys.\ Rev.\ 
Lett.}, {\bf 79}, 4\,297.

\item Mochena, M., and Pollak, M., 1991, {\sl Phys.\ Rev.\ Lett.}, 
{\bf 67}, 109.

\item Ogata, S., and Ichimaru, S., 1989, {\sl Phys.\ Rev.\ Lett.}, 
{\bf 62}, 2\,293.

\item Ortu\~{n}o, M., Talamantes, J., Cuevas, E., and
D\'\i az-S\'anchez, A., 2001, {\sl Phil.\ Mag.\ B}, {\bf 82}, in 
press.

\item P\'erez-Garrido, A., Ortu\~no, M., Cuevas, E., Ruiz, J., and 
Pollak, M., 1997, {\sl Phys.\ Rev.\ B}, {\bf 55}, R8630.

\item P\'erez-Garrido, A., Ortu\~no, M., D\'\i az-S\'anchez, A., and 
Cuevas, E., 1999, {\sl Phys.\ Rev.\ B}, {\bf 59}, 5\,328.

\item Pollak, M., 1970, {\sl Discuss.\ Faraday Soc.}, {\bf 50}, 13.

\item Pollak, M., and Ortu\~{n}o, M., 1985, {\sl Electron-electron 
interactions in disordered systems}, edited by Efros, A.L., and 
Pollak, M., (Amsterdam - Oxford - New York - Tokyo: North Holland), 
p.\ 287.

\item Rujan, P., 1988, {\sl Zeitschr.\ f.\ Physik B (Condensed 
Matter)}, {\bf 73}, 391.

\item Sandow, B., Gloos, K., Rentzsch, R., Ionov, A.N., and 
Schirmacher, W., 2001, {\sl Phys.\ Rev.\ Lett.}, {\bf 86}, 1\,845.

\item Schreiber, M., and Tenelsen, K., 1993, {\sl Europhysics Lett.},
{\bf 21}, 697. 

\item Schreiber, M., Tenelsen, K., and Vojta, T., 1996, {\sl J.\ 
Lumin.}, {\bf 66}, 521.

\item Shklovskii, B.I., and Efros, A.L., 1984, {\sl Electronic 
properties of doped semiconductors}, (Berlin - Heidelberg 
- New York - Tokyo: Springer).

\item Srinivasan, G., 1970, {\sl Phys.\ Rev.\ B}, {\bf 4}, 2581.

\item Swendsen, R.H., and Wang, J.S., 1987, {\sl Phys.\ Rev.\ Lett.},
{\bf 58}, 86.

\item Tenelsen, K., and Schreiber, M., 1994, {\sl Phys.\ Rev.\ B}, 
{\bf 49}, 12\,662.

\item Vaknin, A., Ovadyahu, Z., and Pollak, M., 2000, {\sl Phys.\ 
Rev.\ Lett.}, {\bf 84}, 3\,402.

\item Vojta, T., and Schreiber, M., 1994, {\sl Phys.\ Rev.\ Lett.}, 
{\bf 73}, 2\,933.

\end{description}

\newpage

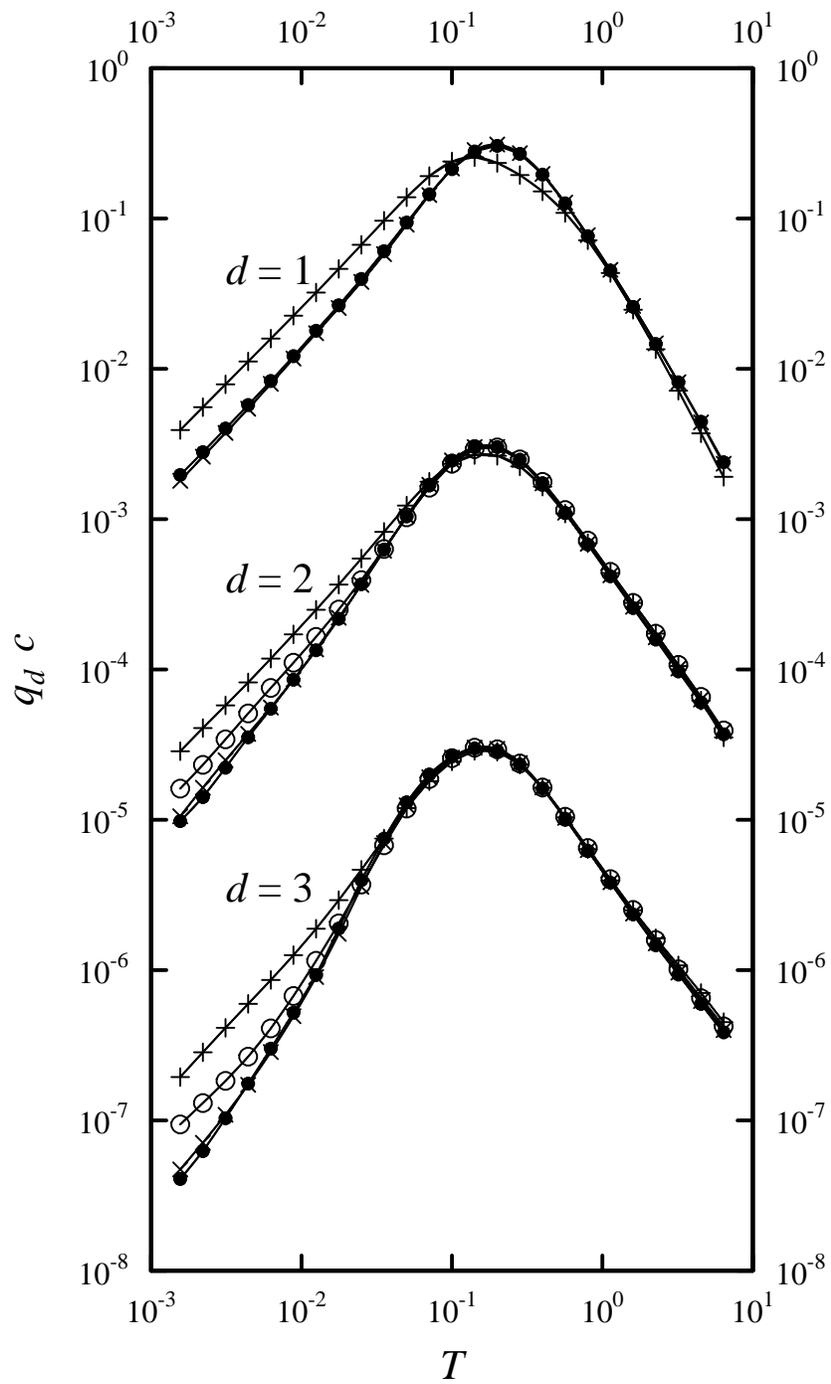
\begin{figure}
\caption{
Specific heat $c$ (heat capacity per site) in dependence of 
temperature $T$ for the lattice model with $B = 1$, $K = 0.5$. For 
the sake of clarity, $c$ has been scaled by a dimension-dependent 
factor: $q_d = 1$, $10^{-2}$, and $10^{-4}$ for $d = 1$, 2, and 3, 
respectively. The symbols denote the the sample sizes as follows. $+$:
3, 9, and 27, $\times$: 12, 100, and 512, $\bullet$: 400, 900, and 
1000 for $d = 1$, 2, and 3, respectively, and $\circ$: 25 and 125, for
$d = 2$ and 3, respectively. All ensemble averages take into account
such a large number of samples that the size of the $1\sigma$-error 
bounds does not exceed the symbol size. E.g., for the largest systems
($\bullet$), we have considered 245, 92, and 50 samples for $d = 1$, 
2, and 3, respectively.
}
\label{fig1}
\end{figure}

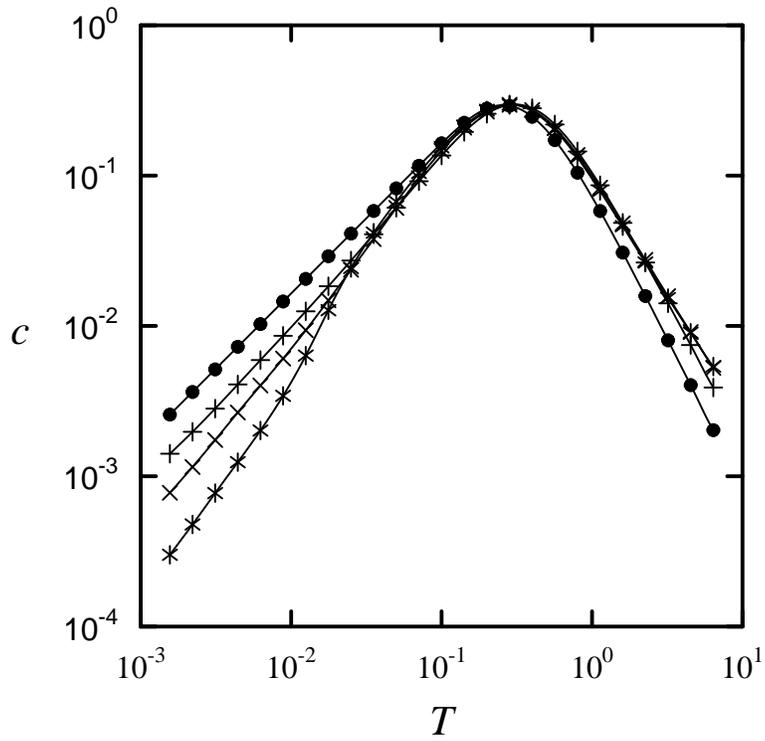
\begin{figure}
\caption{
Dimensional dependence of $c(T)$ for the lattice model with $B = 2$, 
$K = 0.5$. $+$: $d = 1$, $N = 400$, $\times$:  $d = 2$, $N = 900$, 
and $\ast$: $d = 3$, $N = 1000$. For comparison, results for 
interaction switched off, marked by $\bullet$, are included.
}
\label{fig2}
\end{figure}

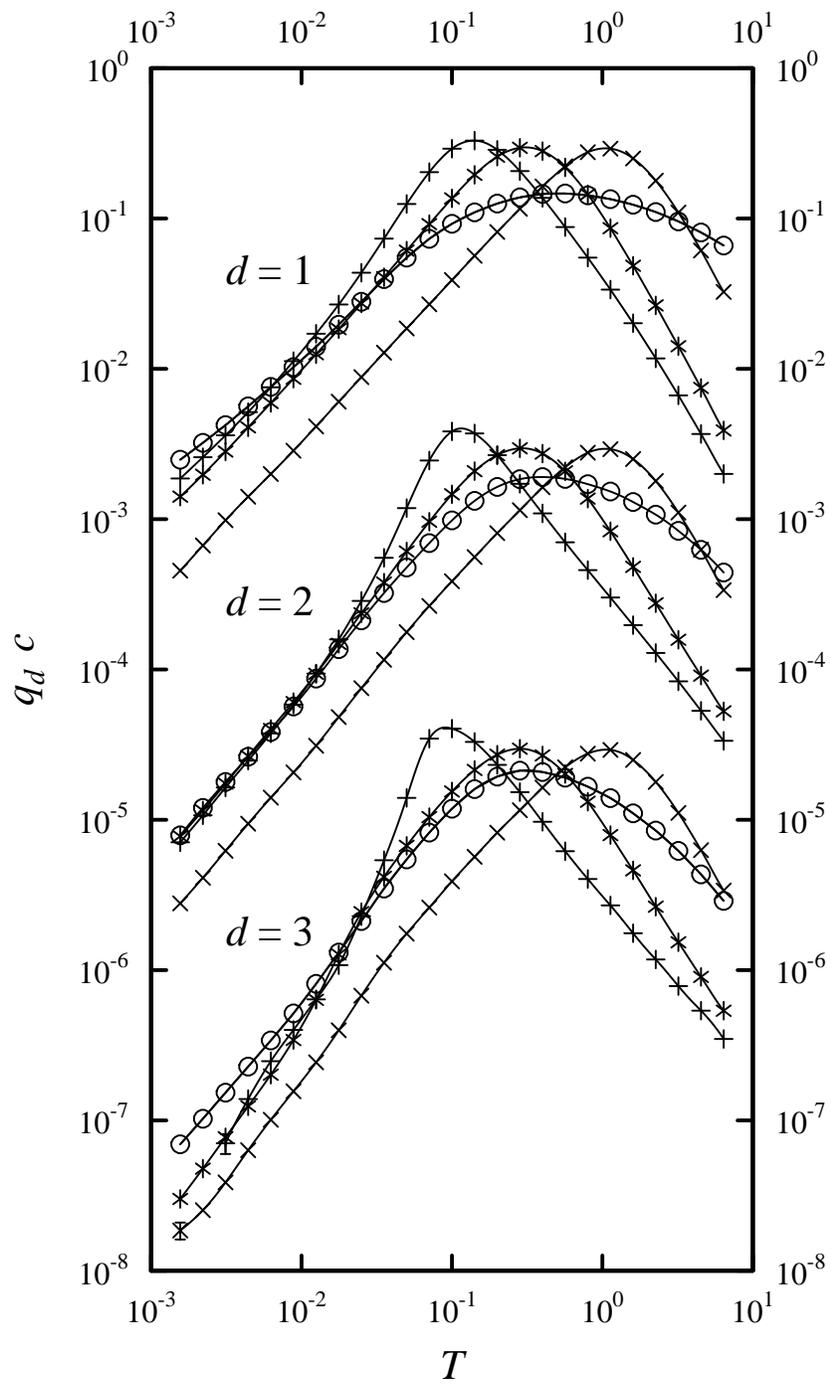
\begin{figure}
\caption{
Influence of the nature and strength of disorder on $c(T)$. For 
clarity, $c$ has been scaled as in Fig.\ 1. $+$: $B = 0.5$, $\ast$: 
$B = 2$, and $\times$: $B = 8$, all for the lattice model; $\circ$: 
random-position model. Error bars denoting the $1\sigma$ deviation are 
only represented if their size exceeds that of the symbol.
}
\label{fig3}
\end{figure}

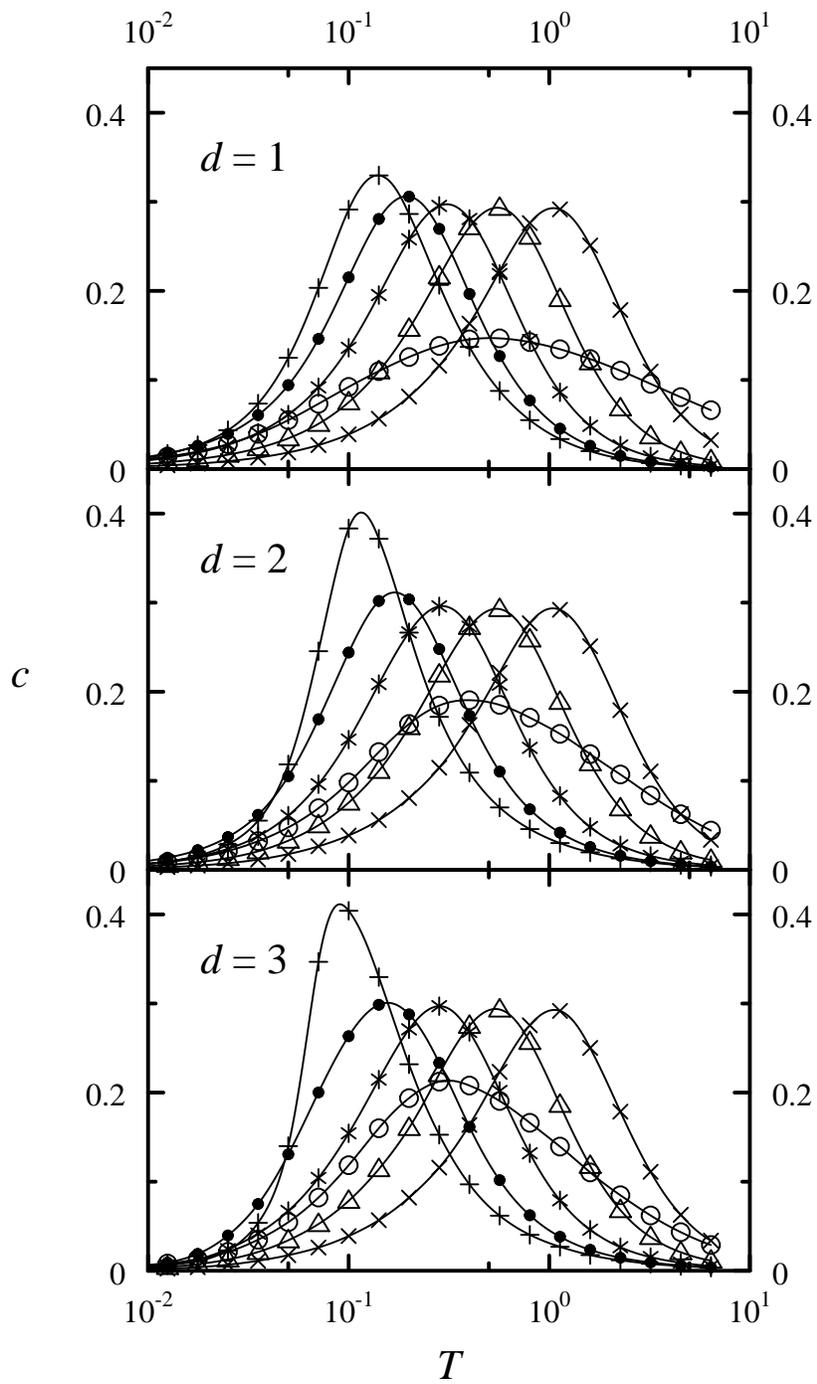
\begin{figure}
\caption{
Influence of the nature and strength of disorder on $c(T)$. $+$: 
$B = 0.5$, $\bullet$: $B = 1$, $\ast$: $B = 2$, $\triangle$: $B = 4$, 
and $\times$: $B = 8$, all for the lattice model; $\circ$: 
random-position model. The spline interpolations are guides to the
eye only.
}
\label{fig4}
\end{figure}

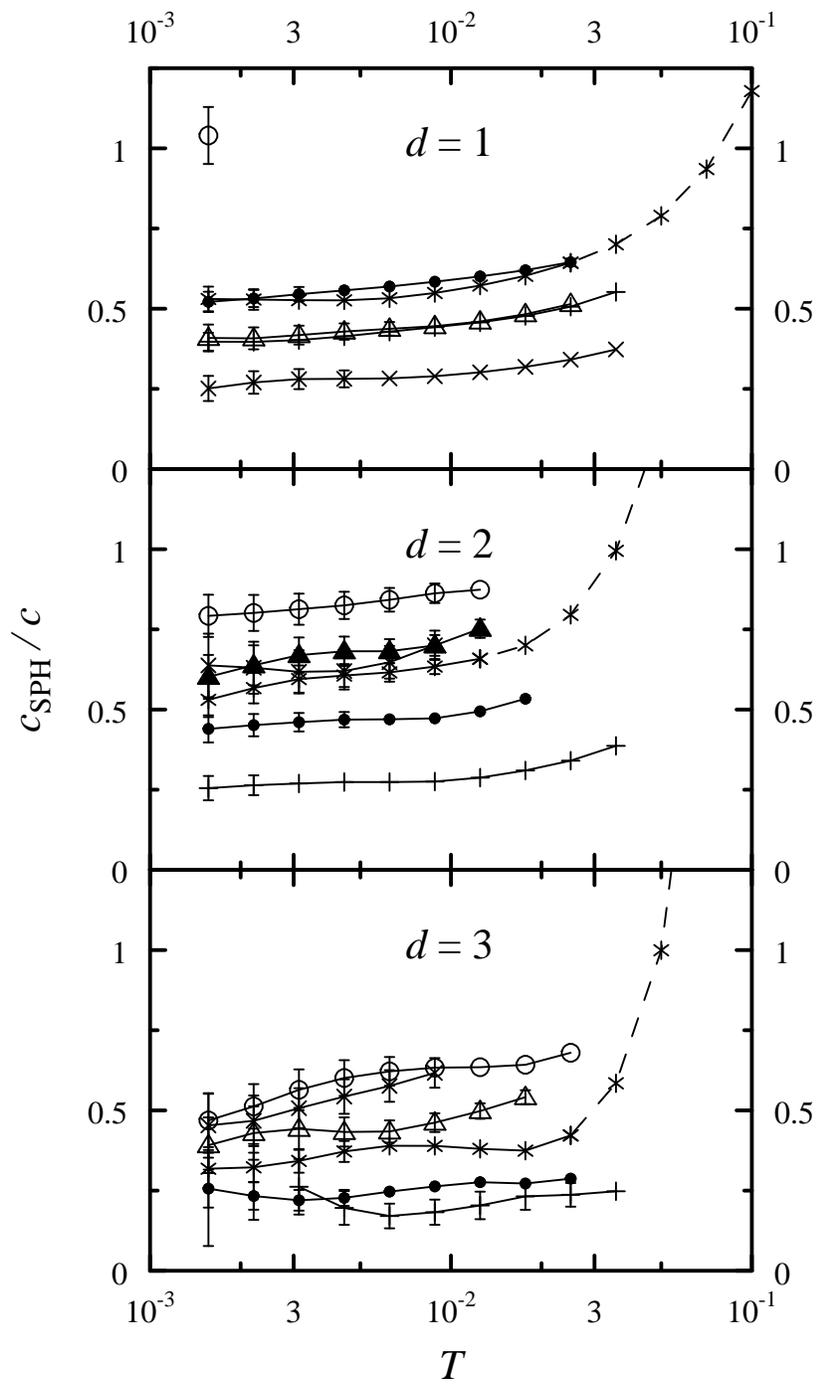
\begin{figure}
\caption{
Comparison of the specific heat value obtained in the single-particle
hop approximation, $c_{\rm SPH}$, with the $c(T)$ values obtained when
the correlations are taken into account completely. For the meaning of
the symbols see Fig.\ 4. The dashed curves represent $c_{\rm SPH}$ 
data in cases, where the overlap of pairs cannot be neglected. For 
$d = 1$, 2 and 3, samples of 24, 400, and 1000 sites, respectively, 
were considered for the lattice model, and samples of 24, 400, and 400
sites were simulated for the random-position model.
}
\label{fig5}
\end{figure}

\end{document}